\documentclass[preprint]{aastex}

\newcommand{\knn}{$k$-NN}
\newcommand{\spitzer}{\emph{Spitzer}}

\slugcomment{To Appear on The Astronomical Journal}

\shorttitle{\knn{} Method for Brown Dwarfs searches}
\shortauthors{Marengo \& Sanchez}

\begin{document}


\title{A \knn{} Method to Classify Rare Astronomical Sources:\\ 
  Photometric Search of Brown Dwarfs with \spitzer{}/IRAC}


\author{Massimo Marengo}
\affil{Harvard-Smithsonian Center for Astrophysics, 60 Garden St,
  Cambridge, MA 02138}
\email{mmarengo@cfa.harvard.edu}

\and

\author{Mayly C. Sanchez\altaffilmark{1}}
\affil{Argonne National Laboratory. 9700 South Cass Avenue.  Argonne,
  IL 60439} 
\altaffiltext{1}{Department of Physics, Harvard University, 17 Oxford
  Street, Cambridge, MA 02138}
\email{mayly.sanchez@anl.gov}




\begin{abstract}

  We present a statistical method for the photometric search of rare
  astronomical sources based on the weighted \knn{} method. A metric
  is defined in a multi-dimensional color-magnitude space based only
  on the photometric properties of template sources and the
  photometric uncertainties of both templates and data, without the
  need to define ad-hoc color and magnitude cuts which could bias the
  search. The metric is defined as a function of two parameters, the
  number of neighbors $k$ and a threshold distance $D_{th}$ that can
  be optimized for maximum selection efficiency and completeness. We
  apply the method to the search of L and T dwarfs in the \spitzer{}
  Extragalactic First Look Survey and the Bo\"otes field of the
  \spitzer{} Shallow Survey, as well as to the search of sub-stellar
  mass companions around nearby stars. With high level of
  completeness, we confirm the absence of late-T dwarfs detected in at
  least two bands in the First Look Survey, and only one in the
  Shallow Survey (previously discovered by \citealt{stern07}). This
  result is in agreement with the expected statistics for late-T
  dwarfs. One L/early-T candidate is found in the First Look Survey,
  and 3 in the Shallow Surveys, currently undergoing follow-up
  spectroscopic verification. Finally, we discuss the potential for
  brown dwarf searches with this method in the \spitzer{} warm mission
  Exploration Science programs.

\end{abstract}







\section{Introduction}\label{sec-intro}

One of the most common problems in astrophysics is the classification
of astronomical sources based on their colors and magnitudes in a given
photometric system. When high spectral resolution data (or a large
number of photometric bands covering the source SED) are available,
this classification, and the extraction of the physical parameters of
the source, is generally achieved by fitting the observations with an
appropriate physical model. In many instances, however, either a
reliable model is not available, or the number of photometric bands is
not sufficient to provide a robust source classification. When data
fitting is not possible, a common fallback solution is to infer the
nature of the sources to be classified by their proximity to ``regions'' in
meaningful color-color and color-magnitude diagrams, 
where the sources of a certain class are expected to be found. 

These regions are in turn defined on the basis of generic physical
considerations (e.g. stars burning H in their cores are located on the
region of the Herzsprung-Russell diagram we call Main Sequence) or by
association with other sources of the same class. A typical
example of this approach in the early years of infrared 
space astronomy were the IRAS color-color diagrams \citep{vanderveen88}
aimed to automatically classify the $2.4 \cdot 10^5$ sources found by
the InfraRed Astronomical Satellite in its bands at 12, 25, 60 and
100~\micron. The diagrams were created by deriving the
IRAS colors of $\sim 5,400$ sources whose nature could
be inferred by the properties of their IRAS Low Resolution Spectra
\citep{iras86}. The resulting diagrams were a grid of polygonal regions
where sources with specific properties (stars, circumstellar envelopes
with varying degrees of optical thickness, planetary nebul\ae{} and
other infrared sources) were expected to be found. As is common in these
cases, the boundaries between the regions were defined arbitrarily by
using a convenient geometrical pattern bisecting known ``template''
sources used for building the diagrams. Most importantly, these
regions did not have an associated statistical meaning, e.g. it was
not possible to quantify how complete and effective was the source
classification provided by these regions.

Other branches of science, however, have developed statistically valid
techniques to attack this kind of unstructured classification
problems, where detailed knowledge of a model is not required, or
not available. The k-Nearest Neighbors (\knn) method \citep{fix51}, in
particular, has been succesfully used as an efficient ``black box''
predictor for problems of pattern recognition and unsupervised machine
learning, in fields ranging from computerized handwriting recognition
\citep{simard93} to automatic classification of satellite imagery
\citep{michie94}, to medical imaging and diagnostics.

In astronomy, k-Nearest Neighbors methods have been traditionally used
to study clustering in the spatial distribution of astronomical
sources (see e.g. \citealt{bahcall81}), by analyzing the statistical
distribution of the distances, on the plane of the sky or in the
3-dimensional space, between each source and its nearest
neighbors. Alternatively, the method has been the base of regression
techniques for parameter fitting (e.g. photometric redshifts, see
\citealt{ball07}).  In this paper, we will instead apply the \knn{}
method in its role of nonparametric classifier, where the class of a
new set of data is decided based on its \emph{distance} from a class
of ``templates'', and where the distance is defined in a
multi-dimensional color and magnitude space. Our implementation of the
method is specifically tuned to the search of rare sources hidden in a
large catalog.

\begin{figure}
\begin{center}
\includegraphics[angle=-90,scale=0.65]{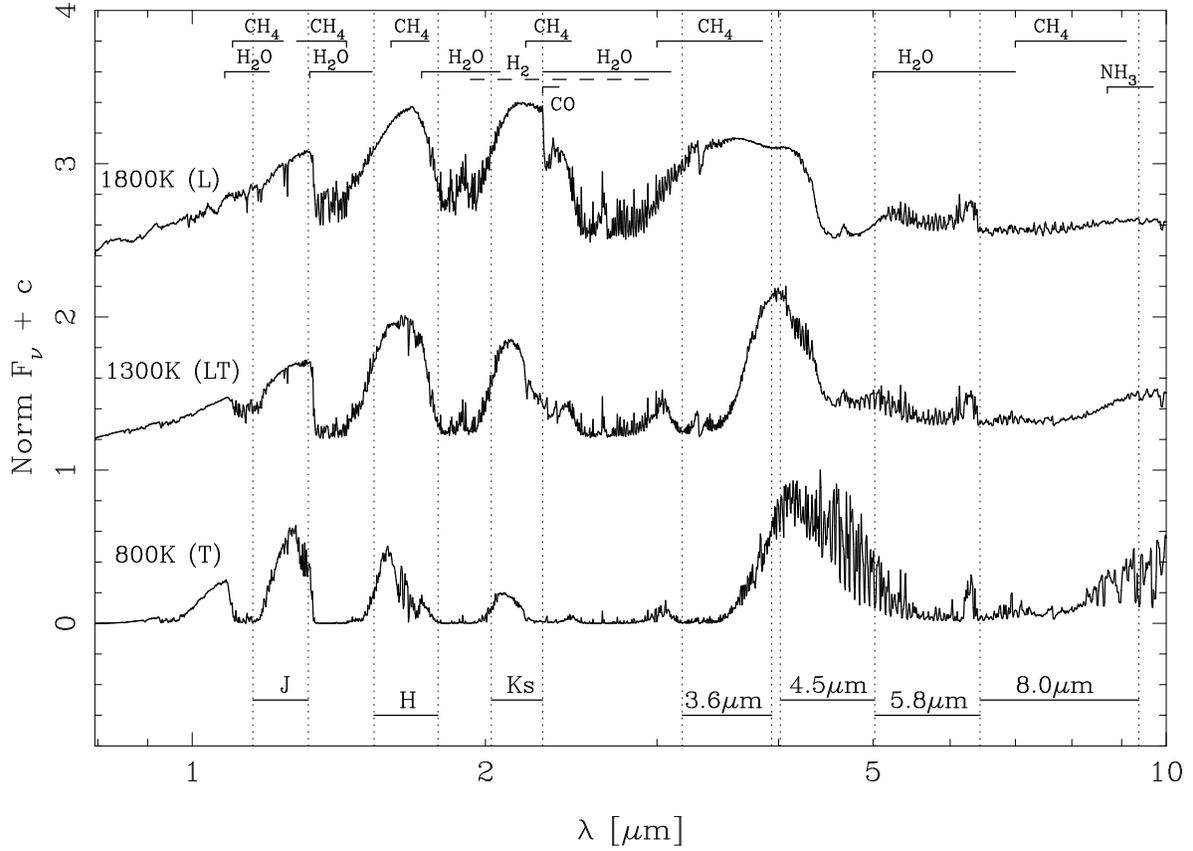}
\end{center}
\caption{Model spectra of brown dwarfs with $T_{eff} \simeq 1800$~K (L
  dwarf), $\simeq 1300$~K (L-T dwarf transition) and $\simeq 800$~K
  (late T dwarf) from \citet{burrows06}. The IRAC and 2MASS
  spectral band-passes are marked, as well as the main molecular
  absorption features in the near- and mid-IR range.}\label{fig-spc}
\end{figure}

To illustrate the effectiveness of the method, we apply our technique
to the search of brown dwarfs with the InfraRed Array Camera (IRAC,
\citealt{fazio04}) onboard the \spitzer{} Space Telescope
\citep{werner04}. As shown by \citet{patten06}, brown dwarfs have
unique colors in the near-IR and IRAC 3.6, 4.5, 5.8 and 8.0~\micron{}
bands, due to the presence of prominent molecular features such as
CH$_4$, H$_2$O, NH$_3$ and CO \citep{oppenheimer98, cushing05,
  roelling04} in the wavelength range covered by the camera (see
Figure~\ref{fig-spc}). These colors provide a powerful discriminant to
identify brown dwarfs within the large photometric catalogs that have
been produced during the \spitzer{} cryogenic mission. The \knn{}
method is particularly suited for this search, because of its high
efficiency in finding ``needles in the haystack'' such as brown
dwarfs, among the galactic general population and the extragalactic
background.

The method is first applied using data from the \spitzer{}
Extragalactic First Look Survey (XFLS, \citealt{lacy05}) and the
\spitzer{} Shallow Survey \citep{eisenhardt04}, which are combined
with ground based optical and near-IR surveys for further refinement
of the candidate sample. The parameter space of the possible color
combinations and \knn{} parameters is explored in order to provide and
quantify the best possible search completeness and
efficiency. Searches using only the two IRAC bands at 3.6 and
4.5~\micron{} are also investigated, to assess the possibility of
brown dwarf detection using only the two channels that will be
available during the post-cryogenic \spitzer{} warm mission.

Section~\ref{sec-description} of the paper describes our
implementation of the \knn{} method, which is then applied in
Section~\ref{sec-bd} to search for field brown dwarfs in the XFLS and
Shallow surveys. In Section~\ref{sec-nstars} the \knn{} method is used
to estimate the efficiency and completeness of IRAC photometric
searches of brown dwarf companions around nearby stars. In
Section~\ref{sec-concl} we summarize the results of these searches,
and discuss other possible applications of the method.


\section{The \knn{} Method}\label{sec-description}

In a typical application of the \knn{} method, as described by
\citet{hechenbichler04}, the class of a test element is selected by a
majority vote of its $k$ nearest templates (where $k$ is optimized to
the specific problem at hand). Template objects for each class, (i.e
the \emph{training sample}) are distributed in the multi-dimensional
space defined by the variables used in the analysis. The class of the
test element is determined by the prevalent class of the $k$ templates
that, according to a given \emph{metric}, are closer to the test
element. Note that in this case the choice of the metric is only
important for the selection of the $k$ nearest templates, after which
the decision is determined by the rule adopted to weight the ``votes''
of these $k$ nearest templates.

The situation is however different in the classical astronomical
setting, where sources of one specific class are searched among a
larger set belonging to many other classes, which are generally not
specified. In this case, a majority vote is not possible and a
different approach is required. In order to address this challenge, we
have developed a novel implementation of the \knn{} method tuned for
the search of rare sources in astronomical photometric catalogs. Our
application requires templates only for the ``search class'', relying
on the assumption that the templates are an accurate representation of
the class, and that the selection variables (colors and magnitudes)
chosen in the analysis are sufficient to provide an effective
discrimination. The only quantitative criterion available for the
selection, and used to determine its statistical validity, will be the
final \knn{} distance of each source in the input catalog, according
to the chosen method. As an example, in section~\ref{sec-bd} we show
the application of our method to the search of brown dwarfs (the
\emph{signal}) in a catalog where most of the other sources are
galaxies or regular stars in different evolutionary stages (the
\emph{background}).

\subsection{The \knn{} Metric}\label{ssec-metric}

\begin{figure}
\begin{center}
\includegraphics[angle=-90,scale=0.55]{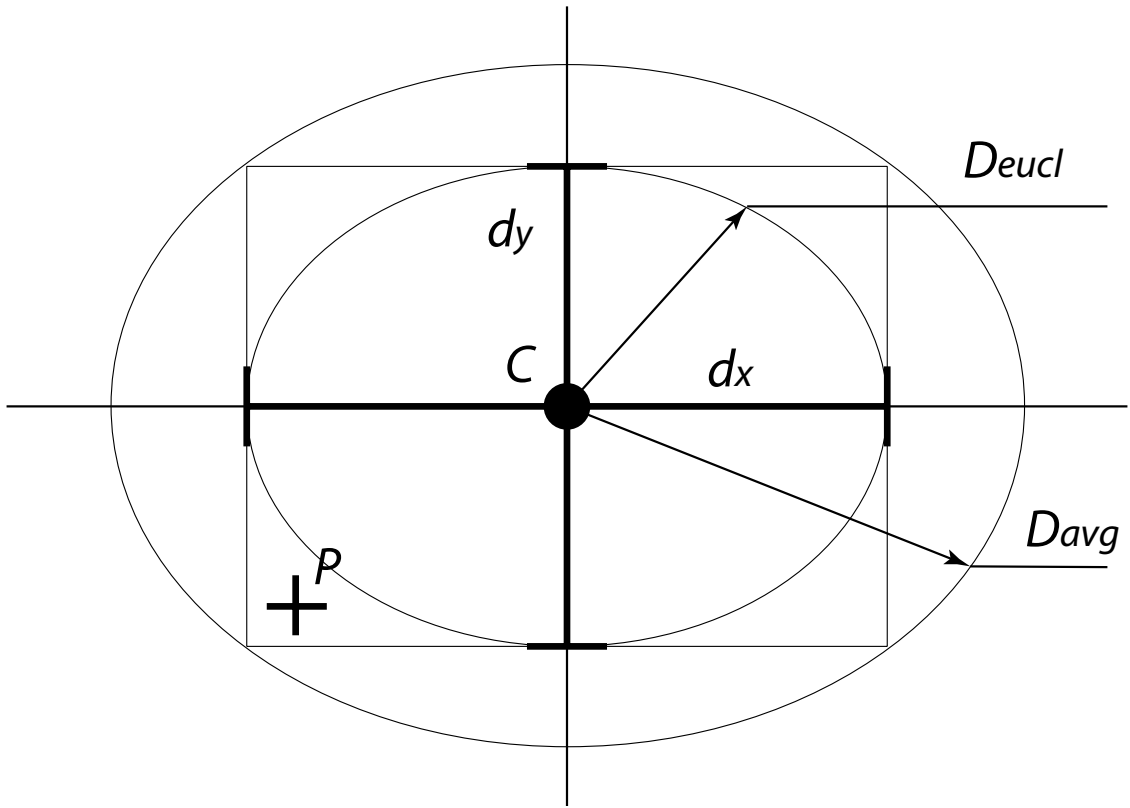}
\end{center}
\caption{Effect of averaging in the metric. Assume that $d_x$ and
  $d_y$ are the distances in the two coordinates $x$ and $y$. If the
  euclidean metric is adopted, the points in the $x,y$ space with
  distance less than $d_x$ and $d_y$ in the two coordinates are
  included in the inner ellipse. If instead the euclidean metric is
  averaged in the two coordinates, any point within the larger ellipse
  will still have distance components less than $d_x$ and
  $d_y$. The point $P$, having individual distances from the center
  $C$ less than $d_x$ and $d_y$ will be excluded by the smaller
  ellipse but still included by the ellipse defined by the euclidean
  average metric.}\label{fig-dist}
\end{figure}

In any \knn{} application the choice of the metric is arbitrary,
and is ultimately determined only on the basis of its ability to
effectively separate the signal from the background. If the analysis
involves $N$ separate variables (in this case colors or absolute
magnitudes), a common choice is the euclidean distance in the
$Nth$-dimensional space, with the individual distances in each
variable averaged (or summed) on the number of dimensions:

\begin{equation}\label{eq-euclid}
D(i,j) = \left\{ \begin{array}{ll}
                 \sqrt{\sum_{l=1}^N d_l(i,j)^2}
                 & \textrm{euclidean distance}\\
                 & \\
                 \sqrt{\sum_{l=1}^N\frac{d_l(i,j)^2}{N}}
                 & \textrm{averaged euclidean distance}\\
                 \end{array} \right.
\end{equation}

\noindent
where $d(i,j)$ is the distance between the source $i$ and the template
$j$ with respect to the variable $l$. The effect of averaging over the
dimensionality $N$ is illustrated in Figure~\ref{fig-dist}: the
average increases the size of the ellipse enclosing test sources
within a given distance from each template. If a source has a unitary
distance from a template along each variable, the distance will be
larger than one in the case of a pure euclidean metric, but still
equal to one in the case of averaged euclidean metric.  The latter is
preferrable for multi-dimensional spaces where is not desirable to
have a metric that tends to become larger as the dimensionality $N$
increases: in other words, it is a good choice to have a distance
close to unity if the individual components of such distance in each
variable are all around unity. For this reason we have adopted the
averaged euclidean distance in this implementation of the \knn{}
method. Furtermore, another advantage of averaging is that it allows
to weight differently the individual components of the metric, in case
some of the variables (colors or magnitudes) have a stronger
discriminatory property for the problem at hand. However, in the
applications presented here we assume for simplicity that all the
variables are equally important, and no weighting is necessary.

To ensure that the distances along each variable play
an equal role in the final metric, a normalization is
required. The metric should take into account, for example, if one
color or magnitude has larger uncertainty than the others. Thus we divide 
each distance by its associated total uncertainty:

\begin{equation}\label{eq-norm}
d_l(i,j) = \frac{x_l(i) - x_l(j)}{\sigma_l(i,j)} \ \ \ ; \ \ \
\sigma_l(i,j) = \sqrt{ \left( \sigma_i^2 + \sigma_j^2 \right) } +
\sigma_s(j) 
\end{equation}

\noindent
where $\sigma_i$ and $\sigma_j$ are the statistical uncertainties of
the data and templates respectively (e.g. their 3$\sigma$ photometric
error) in the variable $l$, and where $\sigma_s$ is a measure of the
non-Gaussian systematic errors of the template $j$ (explained below),
also in the variable $l$.

The final \knn{} distance of a source $i$ is then the (weighted)
average of the distances to the nearest $k$ templates (the optimal
number of neighbors $k$ is determined with the techniques described in
Section~\ref{ssec-apply}):

\begin{equation}\label{eq-knn}
D_{kNN}(i) = \frac{\sum_{j=1}^k D(i,j) \cdot
  w(i,j)}{\sum_{j=1}^k w(i,j)}
\end{equation}

The weights $w(i,j)$ are introduced to reduce the influence of
isolated templates that happen to be much farther away than the other
nearest neighbors. A Gaussian kernel is very effective for this task: 

\begin{equation}\label{eq-w}
w(i,j) = \exp \left[ \frac{- D(i,j)}{k \cdot \sqrt{\sum^N_{l=1}
          \sigma_l(i,j)^2}} \right]^2
\end{equation}

Note that the Gaussian kernel is parametrized with the geometric
average (on all variables $l$) of the same normalization factor
$\sigma(i,j)$ adopted for the individual distances. This is again
necessary to scale the range of the kernel proportionally to the
accuracy of the individual templates. The extra factor $k$ plays the
role of reducing the effectiveness of the kernel as the number of
neighbors increases, which is the intended goal of using a large value
for $k$.

\begin{figure}
\begin{center}
\includegraphics[angle=-90,scale=.55]{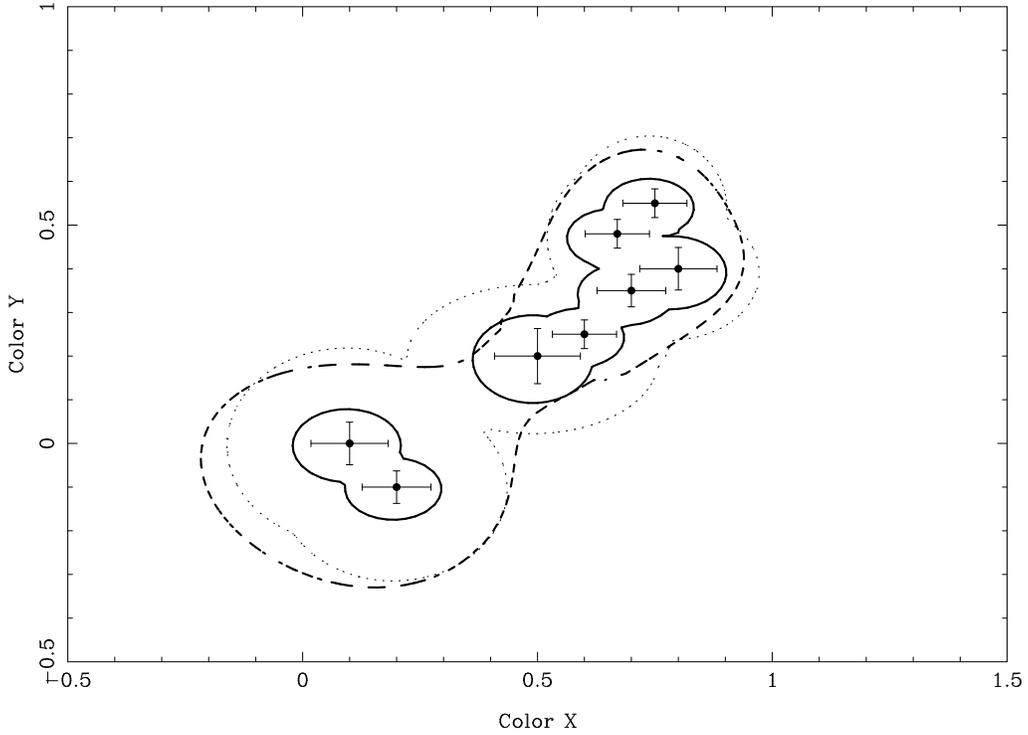}
\end{center}
\caption{\knn{} regions around their templates in a two-dimensional
  color-color diagram. The error bars for each template represent the
  total statistical error in equation~\ref{eq-norm}. The solid line is
  the \knn region for $k = 1$ and no systematic error $\sigma_s(j)$:
  note how the contour encloses the union of the individual regions
  represented in Figure~\ref{fig-dist} for $D_{avg} = 1$. The two
  templates at the bottom are separated enough from the others to form
  a disconnected region. When the sparseness of the templates
  $\sigma_s(j)$ is taken into account, a single region emerges (dotted
  line). For $k = 6$ the region contour becomes smoother, as shown by
  the dashed line. Note that for $k = 6$ the region around the
  isolated sources becomes wider, since $\sigma_s(j)$ is calculated
  reaching templates from the other group.}\label{fig-norm}
\end{figure}

The effect of this normalization on the \knn{} distance is shown in
Figure~\ref{fig-norm}. The contours traced around the template sources
enclose the areas within a given \knn{} distance from the training
sample. The meaning of this region becomes obvious in the case of $k =
1$ (solid line). Thanks to the normalization with the total uncertainty
$\sigma(i,j)$, the region with a radius $D_{kNN} = 1$ is nothing else
than the union of the error ellipses around the templates. A test
source within the region will have a separation from the template
class which is less than the uncertainty in the data and the
templates, and will thus likely be a member of the template class. A
source with $D_{kNN} \ga 1$ will instead have a greater probability of
not being a member, and should be rejected. Note that for $k = 1$ the
border of the region closely follows the location of the templates,
deviating from a smooth line because of the scatter in the
distribution of the training sample. A larger value of $k$, on the
other end, will average on the position of the individual templates,
thus producing a smoother contour, albeit at the risk of excluding
isolated templates from the region. The best value of $k$ 
will be a compromise between these two different tendencies, and needs
to be determined case by case, with the parameter optimization
techniques explained in the following section. 

The solid line in Figure~\ref{fig-norm} was derived without adding the
non-Gaussian systematic error $\sigma_s$ in the templates. As a
result, for any given \knn{} distance, the size of the region is
determined only by the statistical errors in data and templates. A
direct consequence of this, however, is that templates that are
separated by a distance larger than their statistical error will
produce separated regions. This is not desiderable, as in most
astronomical applications the location of the templates is only
partially in control of the astronomer. Due to low statistics it
depends on the chance location, in the color-color and color-magnitude
space, of the template sources that have been possible to
observe. This is particularly true for the search of rare sources,
where only a small number of templates is generally available. The
sparseness of the templates in this case may lead to a serious
problem, as regions of the color-magnitude space where sources of the
template class may exist could be excluded only because no templates
have been observed there at the time the training sample was
assembled. To correct this issue we introduce in our \knn{} metric the
\emph{sparseness factor} $\sigma_s$, which is a measure of how far
apart the templates are with respect to each variable $l$:

\begin{equation}\label{eq-sparse}
\sigma_s(j) = \frac{1}{k} \cdot \sqrt{\sum_{t=1}^k d_l(t,j)^2}
\end{equation}

\noindent
where $\sigma_s(j)$ is defined as the average distance of the template
$j$ from the remaining $k$ nearest templates. The dotted line (for $k = 1$) 
in Figure~\ref{fig-norm} shows how the introduction of the sparseness
factor $\sigma_s$, acknowledging the inadequacy of the template
distribution in a region of the parameter space, is able to reconnect
the region despite the lack of templates between the two sets that are
separated in the solid line region. Particular caution, however,
should be taken while introducing the $\sigma_s(j)$ defined as in
equation~\ref{eq-sparse} in cases where a gap in the templates
distribution is actually expected in the data (e.g. the gap in the
Horizontal Branch for He burning giants in the Herzsprung-Russell
diagram). In such cases, the presence of the gap can only be
noticed, and is statistically significant, when it is sampled by a
large number of templates on both sides of the gap. If
$k$ is chosen to be much smaller than the number of templates in the
gap region, the gap will still be preserved as no template across the
gap will be among the $k$ nearest neighbors in
equation~\ref{eq-sparse}, and the sparseness factor will be smaller
than the width of the gap.


\subsection{Application of the Method}\label{ssec-apply}

The \knn{} metric we have defined in Eq.~\ref{eq-knn} is ideally
suited for selecting rare astronomical sources based on their spectral
properties from a large photometric catalog.

The first step in applying the method is the identification of the
variables to use. These variables can be any combination of colors or
magnitudes. Unless the number of available bands is so small that all
the possible combinations can be tested, the best course of action is
to choose the variables that, based on astrophysical considerations,
provide the best discrimination (e.g. colors sensitive to peculiar
spectral features). Caution must be used to avoid choosing too many
variables: even though one would naively assume that more variables
would produce a better selection, this is not always the case and can
lead to the so-called ``curse of dimensionality'' (see
e.g. \citealt{hastie03}). Also variables that do not carry a
significant discriminating role should be avoided, because they would
dilute the effectiveness of the metric by averaging out more effective
variables. Having multiple variables sensitive to the same physical
property can also be counter-productive, as it biases the metric
towards this one physical characteristic, at the expense of other
equally or more important discriminants. A solution to this issue is
to either to avoid adding variables not contributing any original
discriminant, or to reduce their influence by fine-tuning the \knn{}
metric with appropriate weights.

Once the variables are chosen, the $D_{kNN}$ distance of each source
in the catalog can be determined. The selection is done by comparing
$D_{kNN}$ with a threshold value $D_{th}$ above which the sources are
rejected. For maximum effectiveness the number of neighbors $k$ and
the threshold distance $D_{th}$ have to be optimized for the problem
at hand. The goal of this optimization is to select the smallest
possible number of candidates to follow-up, while preserving the
completeness of the search. In this context, we define the
\emph{completeness} $\cal C$ as the fraction of the objects that are
found, with respect to their expected number. In addition, we define
the \emph{rejection efficiency} $\cal E$ as the fraction of the
background objects that are successfully rejected. An ideal search
will have 100\% completeness (all sources are found) and 100\%
rejection efficiency (all the returned candidates are genuine). In
practice both fractions will be lower than 1, and the search
parameters should be optimized to provide maximum possible rejection
efficiency and completeness.  This optimization can be done using
either the \emph{jackknife} or the \emph{bootstrap} methods
\citep{hastie03}. Both methods attempt to estimate the statistical
distribution of ${\cal C}(k,D_{th})$ and ${\cal E}(k,D_{th})$ to
determine the best values of $k$ and $D_{th}$.

The jackknife method tests the minimum distance for which the
templates are an homogeneous and contiguous set. Given a certain $k$,
one measures the \knn{} distance of each template from the remaining
$n - 1$ (leave-one-out method). Once this is done for all templates,
the completeness is derived as the fraction of templates that are
within any given threshold distance $D_{th}$. While this method is
relatively straightforward, it may be very inaccurate for the search
of rare sources, since the number of available templates is often small
and does not cover uniformly the color and magnitude distribution of
the target objects. Thus we adopt the bootstrap method.

With the bootstrap method we evaluate the completeness by
creating an artificial sample with the characteristics of the
templates. This artificial sample is then tested with the \knn{}
method to estimate how many of these simulated sources are found
within a distance $D_{th}$. The test sample is generated by varying
the template colors and magnitudes, adding a random offset using the
statistical and systematic errors in the templates. The statistical
errors are introduced by adding a Gaussian error equal to the
statistical uncertainty of the templates ($\sigma_j$ in
equation~\ref{eq-norm}). The systematic errors are instead simulated
by adding a random uniform component equal to the amplitude of the
template sparseness factor $\sigma_s$ defined in
equation~\ref{eq-sparse}. With this method it is possible to create an
arbitrary number of simulated signal and background samples of any
size, enabling the study of the statistical distribution of ${\cal
  C}(k,D_{th})$ even when only a small number of templates is
available.

The rejection efficiency $\cal E$ is similarly evaluated with the
bootstrap method. If a ``pure'' background sample can be extracted
from the catalog, then it is just a matter of counting the fraction of
catalog sources that are rejected with any combination of $k$ and
$D_{th}$. In most cases a pure background sample is however not
available but, if the objects in the search class are rare, any small
sub-sample of the full catalog can be assumed to have a very small
contamination of template-class sources. These sub-samples can be
randomized introducing Gaussian noise equal to the data uncertainty
$\sigma_i$ defined in equation~\ref{eq-norm}. Sub-samples that by
chance have a larger level of contamination will appear as outliers,
and can be removed from the final distribution of ${\cal
  E}(k,D_{th})$.

Once the distribution of $\cal C$ and $\cal E$ is known, the
characteristics of the selection problem determine the value of $k$
and $D_{th}$ to choose. If the search needs to be very selective
because a large number of follow-up observations is not affordable,
then the completeness can be sacrificed (generally by using a small
value of $D_{th}$) in favor of high $\cal E$. On the contrary, when
completeness is paramount, a larger threshold distance can be adopted
even though it will result in a large number of candidates. As an
intermediate solution we adopt the value of $D_{th}$ and $k$ for
which ${\cal E}(k,D_{th}) = {\cal C}(k,D_{th})$ is the highest.

After the \knn{} metric is applied with the optimized parameters, all
sources within the threshold distance $D_{th}$ should be considered as
candidates. These candidate sources can be followed-up by applying
further selection criteria (e.g. applying color cuts that could not be
included in the \knn{} metric, or by executing new targeted
observations). If the selected sample is still too big for a follow-up
program, it can be helpful to run the \knn{} method a second time on
the first-run selection, using a different set of variables. This is
particularly effective if there is a concern that some variables have
yielded a lower selection efficiency than expected, due to having been
averaged out in the metric by other variables. In this case it may be
just more efficient to re-apply the \knn{} method using only these
variables, starting from the sources selected in the first run, rather
than trying to improve the efficiency of the first \knn{} run by fine
tuning the weights between the variables.


\section{Searching Brown Dwarfs in \spitzer{} Wide Field
  Surveys}\label{sec-bd} 

As an application of the method, we present the search of brown dwarfs
in \spitzer{} surveys. Brown dwarfs represent the link between main
sequence stars, fully supported by H burning in their cores, and
planets. Given our poor understanding of the lower end in the stellar
mass distribution, an accurate census of the galactic population of
brown dwarfs is of paramount importance to constrain models of star
formation and galactic evolution, and to provide an accurate
measurement of the stellar mass in the Galaxy.

Due to their low luminosity and red colors, brown dwarfs are difficult
to find. The first unambiguous identification of a brown dwarf, Gliese
229B \citep{oppenheimer95}, came only 20 years after the class was
introduced by Jill Tarter in her Ph.D. thesis. In recent years,
however, the availability of deep wide area surveys such as the Two
Micron All Sky Survey (2MASS, \citealt{skrutskie06}), the Deep Near
Infrared Survey of the Southern Sky (DENIS, \citealt{epchtein99}), the
Sloan Digital Sky Survey (SDSS, \citealt{york00}) and the UKIRT (UK
Infrared Telescope) Infrared Deep Sky Survey (UKIDSS,
\citealt{lawrence07}) allowed to identify an increasing number of
brown dwarfs in the solar neighborhood \citep[see
e.g.][]{kirkpatrick00, burgasser02, leggett02, geballe02, burgasser04,
  knapp04, tinney05, chiu06, cruz07, looper07, pinfield08}. However, the total
number of brown dwarfs known to date ($\sim 556$ of the red, dusty L
dwarfs, and $\sim 148$ of the cooler, methane rich T dwarfs, according
to the DwarfArchives.org database) is still too small to
provide a reliable characterization of the substellar mass function.

The sensitivity of the IRAC instrument onboard \spitzer{}, and the
characteristics of its photometric system, raised expectations for a
large increase in the number of brown dwarfs (especially the cooler T
dwarfs) detected using wide area \spitzer{} surveys. These
expectations, however, have not yet materialized. Only three T dwarfs have
been discovered by \spitzer{}: a T4.5 field dwarf in the Extragalactic
\spitzer{} Shallow Survey \citep{stern07}, and two T dwarf companions
around the nearby young stars HN~Peg and HD~3651
\citep{luhman07}. This state of affairs arises from the complexity of
discriminating brown dwarf candidates from the large number of
extragalactic red sources that are within the detection limits of IRAC
observations.

The success of photometric searches ultimately depends on the
efficiency of the selection method required to extract from these
large catalogs a manageable sample of sources for spectroscopic
follow-up. This selection is usually done by applying cuts in the
color and magnitude space (see e.g. \citealt{cruz03} and
\citealt{burgasser03} and references therein). \citet{stern07}, in
particular, used a single IRAC color cut, $[3.6] - [4.5] \geq 0.4$, to
select T dwarfs with deep CH$_4$ absorption in the 3.6~\micron{} IRAC
band, complemented by criteria based on the photometry and the
morphology of the sources in optical bands. These criteria are not
able to discriminate between brown dwarfs and high redshit quasars,
and are limited to the detection of dwarfs of spectral type T3 to
T6. The \knn{} method proposed here has been designed to analyze
datasets in a multi-dimensional color and magnitude space, based only
on the distribution of templates without the need to define a-priori
cuts. Thus it is in principle capable to go beyond these limitations,
opening the search to L and early-T dwarfs and T dwarfs of type later
than T6, without introducing the biases associated with the choice of
the cuts, and provide a more efficient and complete search. This will
be especially important during the \spitzer{} Warm Mission planned
from the spring of 2009 when, upon exhaustion of the cryogenic LHe,
the observatory will be tasked to conduct large area surveys using
only the IRAC bands at 3.6 and 4.5~\micron\ \citep{storrie-lombardi07}.

In this section we study how to improve on the selection efficiency of
\spitzer/IRAC search of brown dwarfs using the \knn{} method. We
explore the \knn{} parameter space to understand the best strategy for
these searches, and what are the requirements, in terms of auxiliary
data, for their success. We also develop techniques that
allow to assess the completeness of the result, necessary to draw
statistically valid conclusions from these searches.

\subsection{Sample Selection}\label{ssec-samples}

As test datasets to illustrate our \knn{} search for brown dwarfs, we
use two publicly available \spitzer/IRAC wide field surveys: the
XFLS \citep{lacy05} and the Bo\"otes field of the IRAC Shallow Survey
\citep{eisenhardt04}.

The XFLS main field covers an area of 3.8~deg$^2$ at high galactic
latitude, observed to a sensitivity reaching a $5\sigma$ Vega
magnitude of 18.9, 18.0, 15.7 and 15.1 at 3.6, 4.5, 5.8 and
8.0~\micron{} respectively (obtained with integration times of at
least 60~sec per pointing). The XFLS main field was chosen for the
availability of extensive auxiliary data, including SDSS and
2MASS. The Bo\"otes field of the Shallow Survey instead covers an area
of 8~deg$^2$ with limiting $5\sigma$ Vega magnitudes of 18.4, 17.7,
15.5 and 14.8 at 3.6, 4.5, 5.8 and 8.0~\micron{}
respectively (integration time $\ge 90$~sec). Deep near-IR $J$ an $K_s$
data has been obtained as part of the FLAMINGO Extragalactic Survey
(FLAMEX, \citealt{elston06}) for 7.1~deg$^2$ of the IRAC field, and in
the optical as part of the NOAO Deep Wide-Field Survey (NDWFS,
\citealt{jannuzi99}).

The main difference between the two samples (apart for the Shallow
Survey being almost twice the area of the XFLS), is in the depth of
the optical and near-IR ancillary data. While SDSS provides $5\sigma$
detection limits of 22.3, 23.3, 23.1, 22.3 and 20.8 in $u'$, $g'$,
$r'$, $i'$ and $z'$ \citep{york00}, NDWFS has $5\sigma$ point source
depths of 27.1, 26.1 and 25.4 in $B_W$, $R$ and $I$ respectively
\citep{stern07}. 2MASS provides $5\sigma$ sensitivity in $J$, $H$ and
$K_s$ of 16.6, 15.9 and 15.1 respectively \citep{skrutskie06}, while
FLAMEX approaches a $5\sigma$ sensitivity of 21.4 and 19.5 in $J$ and
$K_s$ \citep{elston06}. The added depth of the NDWFS and FLAMEX data
provides a powerful tool to resolve ambiguities between red sources in
the IRAC bands of galactic and extragalactic nature. By testing our
method on both datasets we can measure the efficiency of the brown
dwarf search on surveys with different depth and assess the auxiliary
data requirements necessary to enable effective brown dwarf searches
during the \spitzer{} warm mission.

\begin{figure}
\begin{center}
\includegraphics[angle=-90,scale=.65]{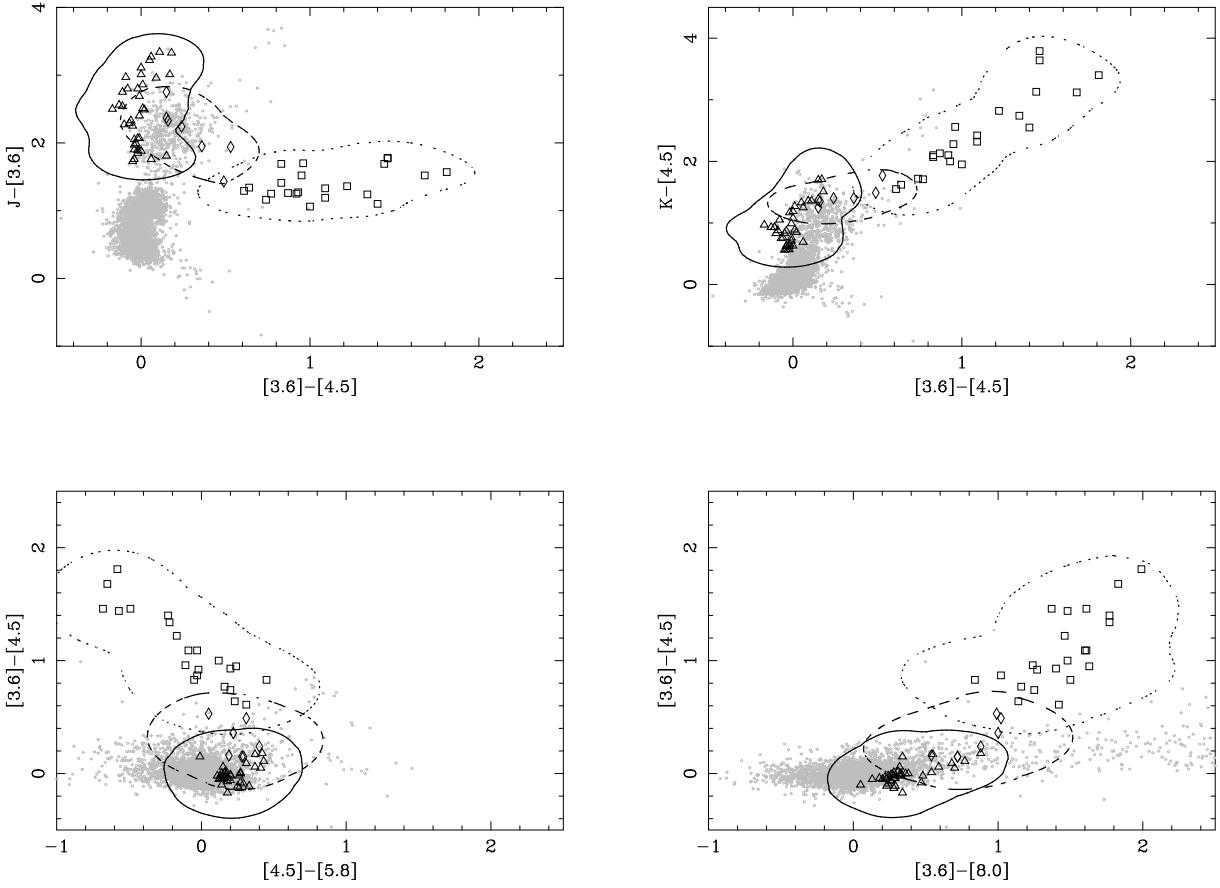}
\end{center}
\caption{Grey points: \spitzer{} First Look Survey sources with good
  photometry ($S/N > 3$) in 6 bands (IRAC bands, plus $J$ and $K$
  from the 2MASS survey). Triangles are \citet{patten06} L dwarfs,
  diamonds are early T dwarfs (spectral type T $<$ 4) and squares are
  late T dwarfs (spectral type T $\ge$ 4). Contours are the \knn{}
  regions defined for $k = 5$ and $D_{kNN} = 1$ for L dwarfs (solid
  line), early T dwarfs (dashed line) and late T dwarfs (dotted
  line).}\label{fig-FLS}
\end{figure}

Figure~\ref{fig-FLS} shows the distribution of point sources from the
XFLS in a number of IRAC and 2MASS colors, compared with the
distribution of 37 L, 7 early-T (T$<$4.5) and 22 late-T (T$>$4.5)
templates from \citet{patten06}. We chose the colors in the diagrams
to maximize the separation between brown dwarfs and other galactic
(clump near zero colors) and extragalactic (long plume with red
colors) sources. In particular (see Figure~\ref{fig-spc}): (1) the
$J-[3.6]$ color is effective in separating the L dwarfs from regular
stars, mainly due to H$_2$O absorption in the $J$ band; (2) the
$K_s-[4.5]$ and $[3.6]-[4.5]$ colors separate the T dwarfs from all
other sources (L dwarfs, regular stars and extragalactic objects), due
to the increasing CH$_4$ absorption in the $K$ and 3.6~\micron{}
bands; (3) the $[4.5]-[5.8]$ color is useful to select again the T
dwarfs, which appear blue because of H$_2$O absorption in the
5.8~\micron{} band; and (4) the $[3.6]-[8.0]$ color is also providing
a strong separation of the T dwarfs due to the methane absorption
which is stronger at shorter wavelength than in the reddest IRAC
band. Color combinations using the $H$ band have a similar
discriminatory power than colors with the $J$ and $K_s$ photometry:
$H-[3.6]$ shows a trend analogous to the $J-[3.6]$ color, and
$H-[4.5]$ has a very similar color trend than $K_s-[4.5]$. The $J-K_s$
colors provide a similar discrimination than the $[3.6]-[4.5]$ colors,
because of CH$_4$ absorption stronger in the $K_s$ than the $J$ band.

The late-T dwarfs appear well separated from any other source, thanks
especially to the $[3.6]-[4.5]$ IRAC color, even though some
contamination persists with red high-redshift quasars having PAH
emission. The early-T and L dwarfs are however more difficult to
discriminate because, once dispersion due to photometric errors is
taken into account, their color space is very similar to the one
occupied by low-redshift galaxies and regular stars. For this reason,
and due to the relative small number of early-T dwarfs in the
\citet{patten06} sample, our \knn{} selection is done for two separate
classes only: one comprising all L and early-T templates, and one with
late-T (with T$>$4.5) dwarfs.

To explore the effect of the presence or absence of individual bands
in the selection efficiency, we have divided our XFLS and Shallow
Survey catalogs in 2 different subsamples: (1) sources having
$3\sigma$ photometry in $J$, $K_s$ and all four IRAC bands; (2)
sources having $3\sigma$ detection in $J$, $K_s$, 3.6 and
4.5~\micron. The first sample is intended to test the effectiveness of
the \knn{} method when all IR bands are available (the $H$ band has
not been considered because of its unavailability in FLAMEX, and
because brown dwarf colors using the $H$ band are very similar to the
colors using $J$ and $K_s$). The second sample is instead designed to
simulate the case of the \spitzer{} Warm Mission, when only the two
short wavelength IRAC channels will be available, and also to avoid
the limitations imposed by the less sensitive 5.8 and 8.0~\micron{}
bands in currently available surveys. For each sample the search is
done using a subset of the available colors, avoiding the repetition
of similar colors, that would dilute the \knn{} metric. The
characteristic of the individual subsamples, their size and the color
combinations used in the \knn{} search are listed in
Table~\ref{tab-samples}. Optical photometry and imaging are not used
at this stage, because only a limited number of the brown dwarfs we
are using as templates have reliable magnitudes at optical
wavelengths. The optical data will however be crucial to refine the
search results later on.


\subsection{Parameter Optimization}\label{ssec-optimization}

The best values of $k$ and $D_{th}$ for the search can be
determined by using the bootstrap method described in
section~\ref{ssec-apply}. The goal is to optimize the \knn{}
parameters in order to have the smallest possible number of candidates
that will need follow-up observations while preserving the
completeness of the search. The jackknife method is not suitable in
this case because of the very small number of available templates in
each brown dwarf class.

The completeness of the search in this case is ${\cal C} = 1 -
n_{FN}/n_{exp}$, where $n_{exp}$ is the number of brown dwarfs
expected to exist in the dataset and $n_{FN}$ is the number of false
negatives (i.e. true brown dwarfs not identified) in the search. The
rejection efficiency can be written as ${\cal E} = 1 - n_{FP}/n_{tot}$
where $n_{FP}/n_{tot}$ is the fraction of false positives, i.e. the
number $n_{FP}$ of incorrectly identified brown dwarfs, with respect
to the total number of sources $n_{tot}$ in the sample.

\begin{figure}
\begin{center}
\includegraphics[angle=0,scale=.70]{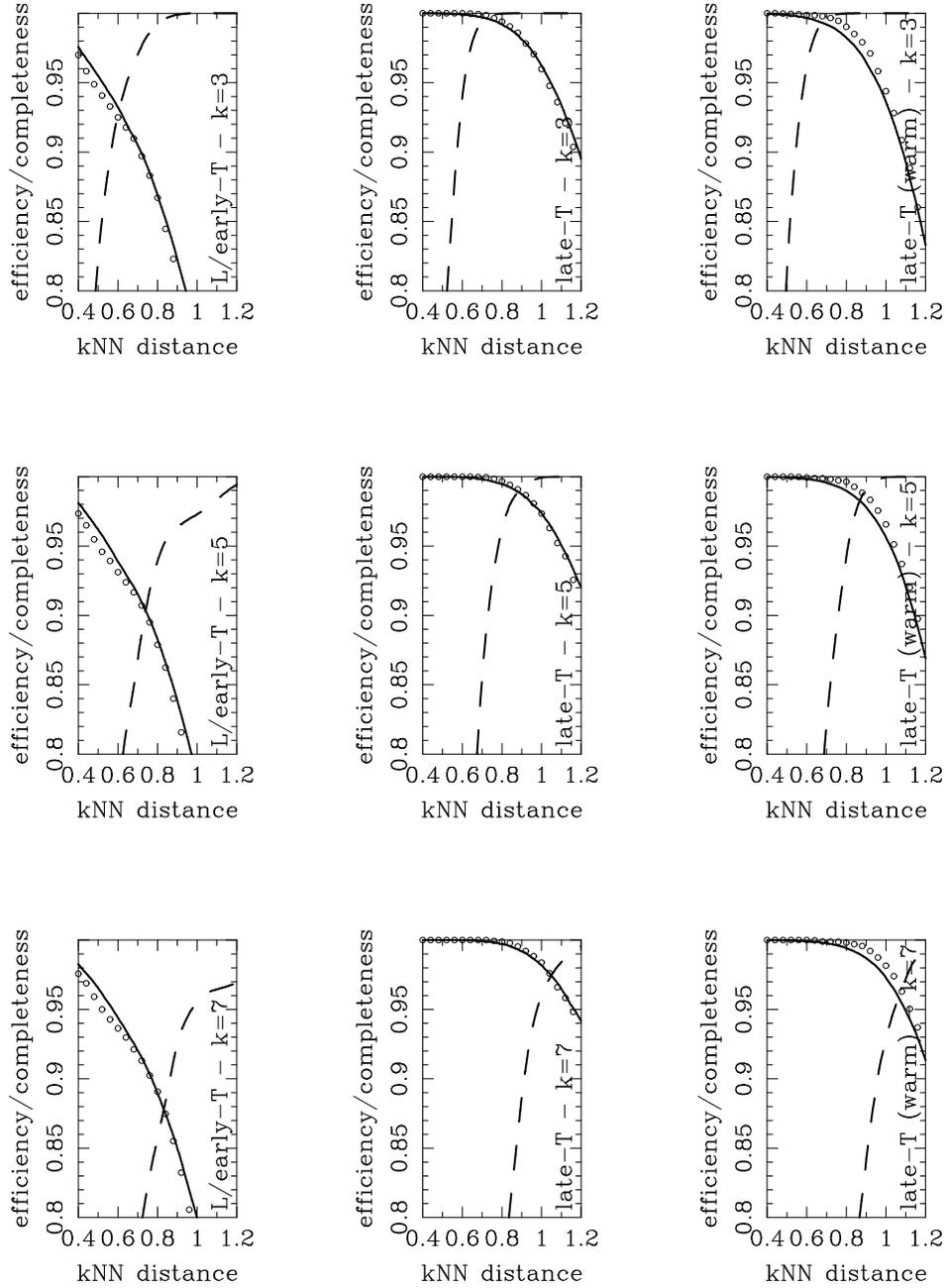}
\end{center}
\caption{Predicted efficiency (solid line) and completeness (dashed
  line) of the L/early-T and late T search as a function of the \knn{}
  distance threshold, for $k = 3$, 5 and 7 (using all colors or IRAC
  only colors). The prediction is based on a Monte Carlo simulation
  seeded with a 20\% subsample of the First Look Survey database. The
  large dots are the actual fractions of sources rejected from the
  full datasets for different values of $D_{th}$.}\label{fig-opt}
\end{figure}

Figure~\ref{fig-opt} shows the rejection efficiency and completeness
for a Monte Carlo test using 100 randomized samples each with 500
background sources and 500 simulated brown dwarfs, for $k = 3$, 5 and
7. The simulations are made for the subsample using IRAC plus $J$ and
$K_s$ bands (for both L/early-T and late-T searches), and for
\spitzer{} warm mission colors (as described in
Table~\ref{tab-samples}). The figure shows the resulting rejection
efficiency and completeness for the XFLS; similar results are obtained
for the Shallow Survey.

The rejection efficiency curve is a decreasing function of $D_{th}$
because when sources with larger \knn{} distances are selected it is
more likely to include false positives in the candidates. On the other
hand, for smaller $D_{th}$ more true brown dwarfs are missed, leading
to a smaller completeness. We have adopted the $D_{th}$ for which the
two curves cross. The values of $D_{th}$ at the $\cal E$ and $\cal C$
crossing point for $K = 3$, 5 and 7 are listed in Table~\ref{tab-ec}
for simulations of the XFLS and Shallow Survey search subsamples. Note
that $D_{th}$ tends to be smaller for L/early-T than for late-T
searches. This is because (as shown in Figure~\ref{fig-opt}) the
efficiency $\cal E$ of L/early-T searches drops faster with $D_{th}$
than in late-T searches, due to the IRAC colors of L/early-T dwarfs
being relatively similar to the colors of regular late spectral type
stars and low redshift galaxies (see Figure~\ref{fig-FLS}): in
L/early-T searches even a small increase in $D_{th}$ results in a
large contamination of background sources and thus in a fast drop of
the efficiency $\cal E$. As a result, in L/early-T searches the
crossing point $\cal E = \cal C$ occurs for smaller values of $D_{th}$
than in late-T searches where, thanks to the unique colors of late-T
dwarfs, the efficiency drops slowly as a function of $D_{th}$.

Table~\ref{tab-ec} shows that the search of late-T dwarfs using IRAC
and near-IR colors combined reaches high level of completeness and
rejection efficiency, up to 99.9\%. Even if only the two short
wavelength IRAC bands are used (as will be the case in the warm
mission), $\cal E$ and $\cal C$ are still similarly high. Note however
that the completeness and rejection efficiency of the warm mission
search tend to be smaller when IRAC photometry is combined with the
deeper FLAMEX dataset, rather then the shallower 2MASS, due to the
higher number of red extragalactic sources cross-correlated with the
IRAC catalog. The L/early-T search is less efficient than the late-T
search, with $\cal C$ and $\cal E$ $\sim 90$\%, due to the higher
contamination of this sample with background sources with similar
colors. 

Rejection efficiency and completeness are generally higher for small
$k$, because in that case the selection region follows more closely
the distribution of the templates. Using small $k$, however, puts us
at risk of depending critically on outliers in the template
class. We adopted $k = 5$ for the searches done using all IRAC
bands. For the search in ``warm mission'' condition, however, we
adopted $k = 3$ to have the maximum possible efficiency, given the
larger size of the initial catalog.


\subsection{\knn{} Search Result and Optical
  Validation}\label{ssec-candidates} 

The results of the \knn{} search for brown dwarfs are presented in
Table~\ref{tab-knn-sel}. The table shows, for each subsample of the
XFLS and Shallow Survey, the number $N_{sel}^{kNN}$ of selected
sources. The actual rejection efficiency ${\cal E'} = 1 -
N_{sel}^{kNN} / N_{tot}$ (where $N_{tot}$ is the number of sources in
each subsample) is also shown. Comparison with Table~\ref{tab-ec}
shows that $\cal E'$ is very similar to the rejection efficiency $\cal
E$ predicted by the bootstrap method Monte Carlo simulations.

Table~\ref{tab-knn-sel} shows that, when all IRAC bands are available,
the \knn{} method alone is efficient enough to reduce the number of
possible candidates to a size small enough to allow visual inspection
of the candidates. As only the \knn{} method is used, the selected
candidates are not biased by the choice of arbitrary cuts, depending
only on a metric with known completeness, normalized on the
statistical uncertainties of both data and templates. Optical data are
however available, and can be used to further reduce the candidate
sample, which is still necessary in the case of L/early-T dwarf
searches, and when requiring only 3.6 and 4.5~$\mu$m photometry
(resulting in a much larger catalog).

The low temperature of brown dwarfs requires their optical colors to
be very red, in contrast with the case of extragalactic objects that
tends to have a flatter optical SED. In particular, according to
\citet{leggett00}, L dwarfs have $i' - z' > 1.6$ in the SDSS
photometric system, while T dwarfs have $i' - z' > 3.0$. According to
\citet{dahn02} L and early-T dwarfs have $R - I > 2.0$ while according
to \citet{stern07} late-T dwarfs must have $R - I > 2.5$ in the NDWFS
bands. These criteria can be applied to all selected sources, as long
as their $i'$ and $z'$ magnitudes (for sources in the SDSS) or $R$ and
$I$ magnitudes (for NDWFS sources) are known. Sources missing optical
photometry can still be considered as brown dwarf candidates, given
that their absence from the optical catalog may be an indication of
very red colors. These colors cannot be introduced directly in the
\knn{} metric because good photometry in the optical bands is missing
for a significant fraction of our templates. By using the optical
bands in the form of ``cuts'' we are introducing a bias associated to
the choice of the cut. However this is still more efficient that doing
the selection using cuts alone, since the candidates have been
pre-selected in an unbiased mode using all the other bands with the
\knn{} metric. This allows us to adopt a less stringent optical color
criteria while still preserving very high efficiency in the
selection. It also allows to retain the candidates selected with the
\knn{} method that are missing optical data (the potentially coolest
late-T candidates), which would be eliminated by default if color cuts
were the only selection criteria.

The application of the criteria described above reduces drastically
the number of viable candidates. The remaining candidates can then be
checked visually from the SDSS and NDWFS plates, to eliminate all
sources that appear extended, that are blended with other sources, or
that are corrupted by artifacts in the images. The remaining
candidates are listed in Table~\ref{tab-XFLS} and
\ref{tab-FLAMEX}. The number of viable candidates after the optical
criteria are applied is listed in Table~\ref{tab-knn-sel}, separately
for the $N^{opt}_{sel}$ sources that possess optical detection, and
the $N^{no \ opts}_{sel}$ that are not detected in the optical
surveys.

Note that when all IRAC, $J$ and $K_s$ photometry is known,
cross-checking with optical data reduces the number of late-T dwarf
candidates to zero for the XFLS, and to only one for the Shallow
Survey. The lone T dwarf selected for the Shallow Survey is in fact
the T4.5 dwarf IRAC J142950.8$+$333011 found by \citet{stern07} in
their search. This result shows that the \knn{} method, when used on
near-IR and \spitzer{} data, in combination with deep optical
photometry is capable to select true late-T dwarfs while rejecting all
other sources of different nature. In fact, the method here described
effectively rejected the second red source found by \citealt{stern07}
color cuts, the $z = 6.12$ radio-loud quasar IRAC
J142738.5$+$331242.

\begin{figure}
\begin{center}
\includegraphics[angle=0,scale=.60]{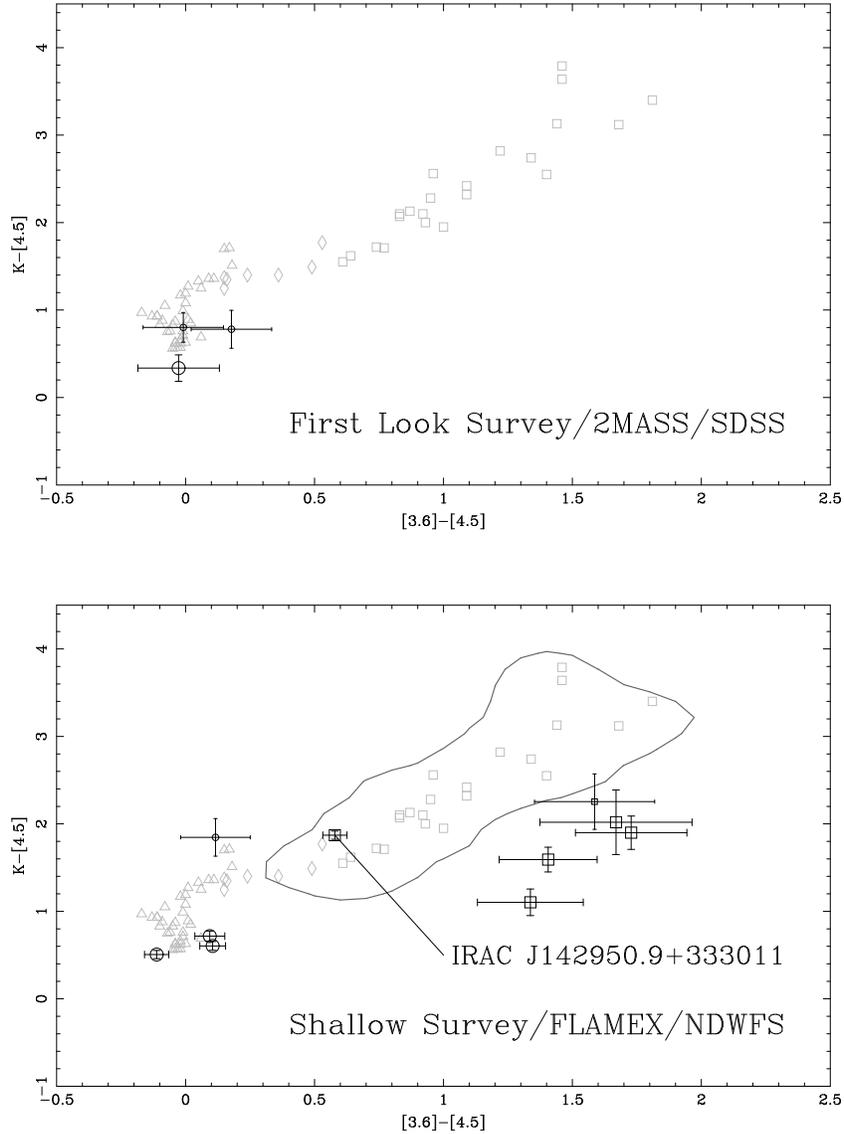}
\end{center}
\caption{Best brown dwarf candidates for the First Look Survey (top)
  and the Shallow Survey/FLAMEX (bottom), selected using $J$, $K$ and
  IRAC colors, satisfying the optical color requirement. Circles are
  L/early-T candidates and squares late-T candidates. Large symbols
  have been verified visually in the SDSS or NDWFS I band images to
  ensure they are single point sources and not extended sources or
  blends. Small symbols are not detected in the optical surveys. Grey
  symbols are the brown dwarf templates. The solid line in the bottom
  panel shows the \knn{} region drawn for $k = 3$ and $D_{th} = 0.8$
  using only the two variables in the plot ($K - [4.5]$ and $[3.6] -
  [4.5]$).}\label{fig-sel-7b}
\end{figure}

Figure~\ref{fig-sel-7b} shows the $K_s - [4.5]$ and $[3.6] - [4.5]$
colors of all the candidates selected in both surveys. Of the plotted
sources, 5 have colors at odd with the brown dwarf templates. These
sources, most likely red background galaxies, have identical colors
than T dwarfs in all bands, with the exception of the two plotted
here. The fact that they have been selected by the \knn{} method is an
example of ``variable dilution'' in the metric, as described in
section~\ref{ssec-apply}. To refine the selection we can apply a
second time the \knn{} method, using only this pair of variables, the
$K - [4.5]$ and $[3.6] -[4.5]$ colors. The region corresponding to $k
= 3$ and $D_{th} = 0.8$ (providing completeness ${\cal C} \ga 99.9$\%
in these two variables) is plotted in Figure~\ref{fig-sel-7b}, and
confirms that these sources are outliers. We flagged them as such in
Table~\ref{tab-FLAMEX}.

After these anomalies are taken into account, we are left with 1
viable L/early-T and no viable late-T candidates in the XFLS (Table~\ref{tab-XFLS}), and 3
viable L/early-T candidates and 1 late-T dwarf (the T4.5 dwarf
J142950.5$+$333011) in the Shallow Survey
(Table~\ref{tab-FLAMEX}). The L/early-T viable candidates need
follow-up observations (currently in progress) to determine
unambiguously their nature.


\section{Searching Low Mass Companions around Nearby
  Stars}\label{sec-nstars} 

As a second example illustrating the case of \knn{} metric using not
only colors but also absolute magnitudes, we show the case of the
search of low mass companions around nearby stars. This search was
conducted as part of the IRAC Guaranteed Time Observations
(P.I. Giovanni Fazio) programs PID~33, 34, 35, 36, and 48
\citep{patten05}. The survey focused on 400 stars within 30~pc from
the Sun, among which all stars and brown dwarfs within 5~pc known to
date. The sample included young stars with age $\leq 120$~Myr, stars
with known radial velocity discovered exoplanets, and the L and T
dwarfs from \citet{patten06} used here as templates. Each star was
imaged to a depth of of $\sim 150$~sec in all IRAC bands, with a field
of view of $\sim 5$~arcmin, sufficient to detect companions at a
distance of $\sim 50$ to 4,000~AU from the primary, to a limiting mass
of $\sim 10$--20~M$_J$. The search has yielded the discovery of two
new T dwarfs from the whole sample, presented in \citet{luhman07}. All
the observations were made in a single epoch, preventing the search of
companions by virtue of their common proper motion with the primary
(except for the few cases in which a deep near-IR image was
available). The candidate selection was instead based on the \knn{}
method described below. While a paper analyzing the search results is
in preparation (Carson et al., 2009), we want here to discuss the
efficiency of the \knn{} method in this particular case of brown dwarf
search.

\begin{figure}
\begin{center}
\includegraphics[angle=-90,scale=.60]{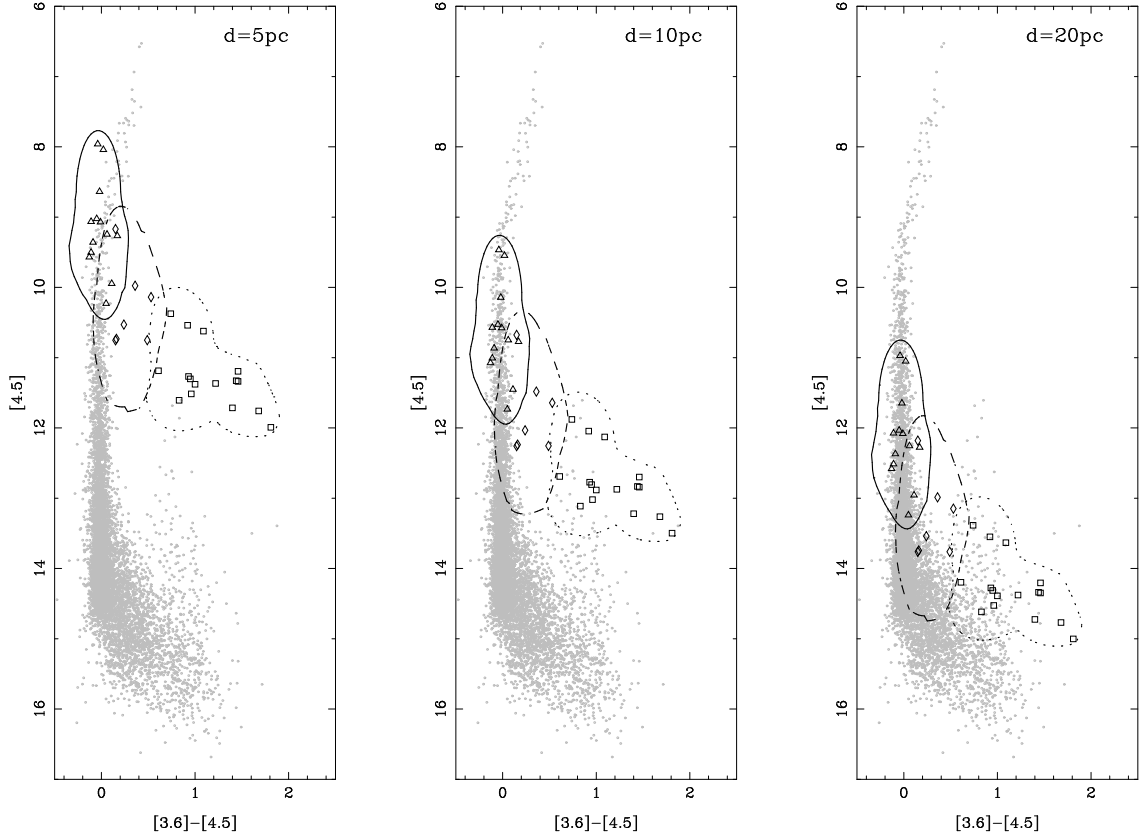}
\end{center}
\caption{Color-magnitude diagram of simulated photometry around a
  nearby star. The data points are from the XFLS. The L, early T and
  late T template [4.5] magnitudes are computed for brown dwarfs at a
  distance of 5 (left), 10 (center) and 20~pc (right)
  respectively. The regions are plotted for $k = 5$. Symbols are the
  same as in previous plots.}\label{fig-mag-ns} 
\end{figure}

Figure~\ref{fig-mag-ns} shows simulated $[4.5]$ vs. $[3.6]-[4.5]$
photometry of L, early-T and late-T dwarfs around a nearby star at 5,
10 and 20~pc. To simulate the photometry of background sources around
the primary star, we have used the XFLS catalog, which is an adequate
representation of a background field projected at high galactic
latitude. For nearby stars projected closer to the plane of the
Galaxy, the proportion of extragalactic/galactic sources will be
smaller, reducing the contamination from red extragalactic sources,
while contamination from red galactic sources is likely to be
increased. The latter (mass losing evolved stars and young stellar
objects) may however be discriminated by means of auxiliary infrared
observations capable of detecting the thermal emission from their
circumstellar dust. Figure~\ref{fig-mag-ns} also shows \knn{} regions
plotted for $k = 5$ and $D_{th} = 1$. By adding the $[4.5]$ magnitude
to the \knn{} variables the selection is in principle improved because
a large fraction of the red background extragalactic sources are
fainter than the expected brightness of brown dwarf companions. This
is especially true for T dwarfs around stars within 5--10~pc from the
Sun. This advantage is reduced for searches around farther stars,
since the brightness of T dwarf companions at $d \simeq 20$~pc is the
same of the extragalactic background.

We have estimated the rejection efficiency and completeness of this
search using the same bootstrap method described in
section~\ref{ssec-apply}. Table~\ref{tab-ec-ns} shows that
for early-T dwarfs the selection efficiency reaches very high values (up
to 99\% for primaries at 5~pc). For late-T dwarfs the rejection
efficiency is approximately the same that is obtained by using only
colors as selection criteria, and the inclusion of absolute magnitudes
does not result in a dramatic improvement in the search
effectiveness (suggesting that the \knn{} color selection is already
close to maximum efficiency). The $\cal E$ and $\cal C$ obtained in
the simulations, are adequate for this kind of search: the typical
number of sources in the $\sim 400$ stars part of the IRAC companion
search program had an average of $\sim 100$ background sources
each. With this efficiency, for each field the chosen \knn{} criteria
selected up to 3 candidates, many of them actually present in at
least one 2MASS map, and could be ruled out either by the absence of
proper motion, or because they did not possess the correct near-IR
colors. The few remaining candidates were followed-up
spectroscopically and by acquiring deep near-IR images, resulting in
the two new T dwarfs found around HN~Peg and HD~3651, presented in
\citet{luhman07}.


\section{Discussion and Conclusions}\label{sec-concl}

Based on the photometry given by \citet{patten06}, the sensitivity of
the XFLS and the Shallow survey in the IRAC bands allows for the
detection of late-T dwarfs (T4 to T8 spectral type) up to a distance
of $\sim 20$~pc. This limit is determined mainly by the lower
sensitivity of the 5.8 and 8.0~\micron{} bands, for the latest
spectral types. If only the 3.6 and 4.5~\micron{} bands are used, as
in the \spitzer{} warm mission, the higher sensitivity allows
detection of late-T dwarfs up to a distance of $\sim 50$~pc. The
detection limit at the L-T boundary is $\sim 60$~pc ($\sim 140$~pc if
only the 3.6 and 4.5~\micron{} bands are required). Within this volume
(corrected for the Malmquist bias), we can expect a brown dwarf search
to be as complete as ${\cal C} \ga 98$\%, as estimated in
section~\ref{ssec-optimization} (multiplied by the completeness of the
original survey, and corrected for binarity).

Searches using 2MASS photometry, however, will have more stringent
limits, of $\la 5$~pc for late-T dwarfs and $\sim 25$~pc at the L-T
boundary. The sensitivity of the FLAMEX survey is such that any L or T
dwarf detected in the IRAC 3.6 and 4.5~\micron{} bands will also be
detected in the $J$ band, even though the lower sensitivity of the $K$
band limits the maximum detection distance for a T8 dwarf to $\sim
32$~pc. The very small number of brown dwarf candidates that are not
optically detected in at least the $I$ band shows that the depth of
the NDWFS is not a significant constraint in the search of brown
dwarfs. The main limitation rather comes from the depth of the IRAC
data. 

These considerations come into play to understand the potential for
brown dwarf searches in the recently approved Exploration Science
surveys in the \spitzer{} warm mission. The requirements for brown
dwarf detection are clearly a large survey area, deep observations and
the availability of matching near-IR and possibly optical data. Of the
approved warm mission programs, three satisfy these requirements: the
``\spitzer{} Extragalactic Representative Volume Survey (SERVS)'' (PI
Mark Lacy, PID 60024), the ``\spitzer{} Extended Deep Survey (SEDS)''
(PI Giovanni Fazio, PID 60022) and the ``GLIMPSE360: Completing the
\spitzer{} Galactic Plane Survey'' (PI Barbara Whitney, PID 60020).

Our analysis shows that when near-IR data of sufficient depth are
available (as in the case of the FLAMEX survey), the search for late-T
dwarfs using the photometric \knn{} technique described in
section~\ref{sec-bd} is extremely efficient and complete (more than
99.8\% $\cal E$ and $\cal C$). Once a single $I - R$ or $i' - z'$
optical color is applied, the final selection produces a very small
number of viable candidates to be checked individually (4 L/early-T
and 1 late-T viable candidates). It is worth noting that the only
late-T dwarf candidate selected in our search is indeed a T4.5 dwarf,
as discovered by \citet{stern07}. With ${\cal C} \simeq 98$\%
completeness we can assert that this is the only late-T dwarf present
in the volume of the survey (7.1~deg$^2$ for the FLAMEX field with a
depth of $\sim 32$~pc), even when only the two short wavelength IRAC
bands are considered. This number is consistent with the results from
the T dwarf UKIDSS DR2 Large Area Survey \citep{pinfield08}, that
estimates $17 \pm 4$ late-T dwarfs in an area of 280~deg$^2$ for a
depth of $K \simeq 18.2$ (corresponding to a maximum distance of $\sim
18$~pc for T8 dwarfs). By factorizing the search volume between the
two surveys\footnote{$V_2/V_1 \propto \Omega_2/\Omega_1 \cdot \left(
    d_2/d_1 \right)^3$, where $\Omega_{1,2}$ are the survey areas and
  $d_{1,2}$ their search distance limit} one predicts $\sim 2$--3
late-T dwarfs in the volume of the Shallow Survey analyzed in this
paper. This is consistent with our result (1 viable candidate in the
FLAMEX search), upon considering Poisson statistics. The number of
L/early-T candidates that we found shows that, with ${\cal C} \sim
85$\% completeness, $\le 3$ L/early-T dwarfs are present in the
Shallow Survey searchable volume (8~deg$^2$ $\times$ 60~pc).

These numbers can be used to estimate the potential yield of brown
dwarfs in the approved warm \spitzer{} Exploration Science
surveys. The most promising \spitzer{} warm mission project for the
application of the \knn{} method here described is the SERVS
survey. With a total area of 18~deg$^2$, and a total exposure time of
600~sec per pointing, it will probe a brown dwarf volume almost 9
times larger than the IRAC Shallow Survey (search depth of $\sim
80$~pc for late-T dwarfs\footnote{Search depth $d$ scales as the
  limiting flux $F^{1/2}$, in turn scaling as the signal-to-noise
  ratio S/N $\sim t^{1/2}$, where $t$ is the exposure time. This gives
  $d \sim t^{1/4}$ and thus $V_2/V_1 \propto \Omega_2/\Omega_1 \cdot
  \left( t_2/t_1 \right)^{3/4}$}). A large fraction of the survey will
overlap with the VIDEO VISTA
survey\footnote{http://www.vista.ac.uk/index.html}, providing a depth
of 25.7, 24.6, 24.5, 24.0 and 23.5~mags in the $z$, $Y$, $J$, $H$ and
$K$ bands, more than matching the depth of SERVS in the IRAC 3.6 and
4.5~\micron{} bands. Based on these sensitivity, SERVS may find as
many as $\sim 27$ late-T dwarfs and a large number of L and early-T
dwarfs.

The SEDS program, on the other hand, has a much smaller survey area
($\sim 0.9$~deg$^2$) but a much longer integration time (12~hours per
pointing). This scales down to a searchable volume of $\sim 12$ times
the total volume of the Shallow Survey (search depth of $\sim 230$~pc
for late-T dwarfs). This survey can potentially provide as many as
$\sim 36$ late-T dwarfs, even though a decrease in the brown dwarf
density should be expected as the survey probes farther distances from
the galactic mid-plane. This survey, however, may be limited by the
challenge of finding ancillary optical and near-IR data matching the
depth of the IRAC photometry. While this may limit the effective
late-T dwarf searchable volume, the potential search depth offered by
the IRAC data, in a high galactic latitude region, will provide an
important test for the vertical distribution of the brown dwarf
population in the Galaxy.

The GLIMPSE360 survey, instead, compensates the rather shallow
coverage (36~sec integration time for each pointing) with a very large
survey area (187~deg$^2$). A substantial portion of this area is also
covered by the UKIDSS survey \citep{lawrence07}, implying that $\sim
11$ of the 17 late-T dwarfs estimated for the whole UKIDSS may be
present in the GLIMPSE360 area. According to the GLIMPSE360
consortium, more detailed simulations by \citet{burgasser04} predict
70~T0, $\sim 100$~T5 and $\sim 15$--20~T8 dwarfs in the survey search
area. The challenge will be to isolate these brown dwarfs from the
high-confusion galactic field, and distinguish them from other red
galactic sources. The \knn{} method can play an important role for
this task.

The main advantage of the \knn{} method presented in this paper is
that it allows to perform photometric searches using a large number of
color and magnitude variables, defining complex regions that closely
follow the distribution of the sources to be selected. While similar
regions can be defined manually, our method prevents the introduction
of biases in the selection due to the choice of the cuts. With the
\knn{} method the selection regions in the color/magnitude space are
only based on the photometric properties of the search class and the
statistical uncertainties of the data. Also, the \knn{} search can be
controlled by just two parameters (the number of neighbors $k$ and the
threshold distance $D_{th}$), instead of many arbitrary cuts, which
allows to quickly experiment different combination of variables, and
optimize the search for maximum rejection efficiency and completeness.

The examples presented in this paper show that the \knn{}
method is an effective procedure for the search of field and companion
brown dwarfs in \spitzer{} wide field surveys. This can be an
important asset for the \spitzer{} warm mission surveys. As these
surveys will image areas of the sky where deep photometric catalogs in
the optical and near-IR are already available, or in progress, the
\knn{} method can effectively select T dwarf candidates, potentially
leading to a significant increase in the known number of members in
this class. This is by no means the only potential application for
this method. The method is general enough to allow the photometric
selection of sources of any kind, as long as a sample of templates is
available. If enough classes of templates are used, the \knn{}
method can be the engine for the photometric classification of all
sources in a survey, by attributing to each source the class with the
highest \knn{} score. We are currently applying this method of
classification to the point source catalog of the SAGE survey
\citep{meixner06}.




\acknowledgments

This work is based in part on observations made with the \spitzer{} Space
Telescope, which is operated by the Jet Propulsion Laboratory,
California Institute of Technology under a contract with NASA. This
publication makes use of data products from the Two Micron All Sky
Survey, which is a joint project of the University of Massachusetts
and the Infrared Processing and Analysis Center/California Institute
of Technology, funded by the National Aeronautics and Space
Administration and the National Science Foundation. It also used data
from the Sloan Digital Sky Survey (see full acknowledgment at
http://www.sdss.org/collaboration/credits.html), and software provided
by the US National Virtual Observatory, which is sponsored by the
National Science Foundation. We thank the \spitzer{} Shallow Survey
and FLAMINGO Extragalactic Survey (FLAMEX) teams for permission to use
data from their surveys. We also thank the National Optical Astronomy
Observatory (NOAO) Deep Wide-Field Survey Team for providing the
optical and near-IR imaging data used in the Bo\"otes field. NOAO is
operated by the Association of Universities for research in Astronomy
(AURA), Inc., under a cooperative agreement with the National Science
Foundation. The authors would finally like to thank Peter Eisenhardt,
Daniel Stern, Mark Brodwin and Buell Jannuzi for useful discussions and
suggestions, and the anonymous referee for insightful comments that
helped improving this manuscript.




{\it Facilities:} \facility{Spitzer (IRAC)}, \facility{2MASS},
\facility{SDSS}, \facility{NVO}

\clearpage







\begin{deluxetable}{lrl}
\tablecaption{Sample selection\label{tab-samples}}
\rotate
\tablewidth{0pt}
\tablehead{ \colhead{Sample} & \colhead{$N_{tot}$} &
  \colhead{\knn{} Colors Used} }
\startdata
XFLS IRAC/2MASS & 4,552
   & $[3.6]-[4.5]$, $[3.6]-[8.0]$, $[4.5]-[5.8]$, $J-[3.6]$, $K_s-[4.5]$ \\
XFLS IRAC Warm & 8,133
   & $[3.6]-[4.5]$, $J-[3.6]$, $K_s-[4.5]$, $J-K_s$ \\
Shallow Survey IRAC/FLAMEX & 15,847
   & $[3.6]-[4.5]$, $[3.6]-[8.0]$, $[4.5]-[5.8]$, $J-[3.6]$, $K_s-[4.5]$ \\
Shallow Survey IRAC Warm & 71,590
   & $[3.6]-[4.5]$, $J-[3.6]$, $K_s-[4.5]$, $J-K_s$ \\
\enddata
\end{deluxetable}
\clearpage

\begin{deluxetable}{lccccccccc}
\tabletypesize{\footnotesize}
\tablecaption{\knn{} efficiency and completeness optimization\label{tab-ec}} 
\tablewidth{0pt}
\tablehead{ 
& \hspace{0.3cm} & \multicolumn{2}{c}{L/early-T} &
  \hspace{0.3cm} & \multicolumn{2}{c}{late-T} &
  \hspace{0.3cm} & \multicolumn{2}{c}{late-T (warm mission)}\\
& & \colhead{$D_{th}$} & \colhead{$\cal C = \cal E$} &
  & \colhead{$D_{th}$} & \colhead{$\cal C = \cal E$} &
  & \colhead{$D_{th}$} & \colhead{$\cal C = \cal E$} }
\startdata
\sidehead{First Look Survey}
$k = 3$ && 0.62 & 93.1\% && 0.72 & 99.7\% && 0.68 & 99.3\% \\
$k = 5$ && 0.74 & 90.3\% && 0.89 & 98.9\% && 0.87 & 98.3\% \\
$k = 7$ && 0.83 & 87.7\% && 1.05 & 97.5\% && 1.06 & 95.9\% \\
\\
\hline
\sidehead{Shallow Survey \& FLAMEX}
$k = 3$ && 0.56 & 89.7\% && 0.76 & 99.9\% && 0.63 & 97.8\% \\
$k = 5$ && 0.68 & 85.2\% && 0.97 & 99.8\% && 0.81 & 95.3\% \\
$k = 7$ && 0.76 & 83.3\% && 1.17 & 99.4\% && 0.99 & 92.2\% \\
\enddata
\end{deluxetable}
\clearpage

\begin{deluxetable}{lrrrrrrr}
\tablecaption{\knn{} search result\label{tab-knn-sel}} 
\tablewidth{0pt}
\tablehead{ & \colhead{$k$} & \colhead{$D_{th}$} & \colhead{$N_{tot}$} &
               \colhead{N$_{sel}^{kNN}$} & \colhead{$\cal E'$} &
               \colhead{N$_{sel}^{opt}$} & \colhead{N$_{sel}^{no \, opt}$} }
\startdata
\sidehead{First Look Survey}
L/early-T          & 5 & 0.74 &  4,552 &   455 & 90.0\% &  1 &  2 \\
late-T             & 5 & 0.89 &  4,552 &    45 & 99.0\% &  0 &  0 \\
late-T (warm)      & 3 & 0.68 &  8,133 &    17 & 99.8\% &  0 &  0 \\
\\
\hline
\sidehead{Shallow Survey \& FLAMEX}
L/early-T          & 5 & 0.68 & 15,847 & 2,831 & 82.1\% &  3 &  1 \\
late-T             & 5 & 0.97 & 15,847 &    40 & 99.7\% &  1 &  0 \\
late-T (warm)      & 3 & 0.63 & 71,590 & 1,582 & 97.8\% &  4 &  1 \\
\enddata
\end{deluxetable}
\clearpage

\begin{deluxetable}{rllrrrrrrrll}
\tabletypesize{\footnotesize}
\rotate
\tablecaption{XFLS brown dwarf candidates\label{tab-XFLS}}
\tablewidth{0pt}
\tablehead{\colhead{\#} & \colhead{RA} & \colhead{Dec} &
           \colhead{J} & \colhead{H} & \colhead{K} &
           \colhead{[3.6]} & \colhead{[4.5]} & \colhead{[5.8]} &
           \colhead{[8.0]} & \colhead{Type} & \colhead{Notes} }
\startdata
1 & 260.38831 & $+$59.27060 &  16.36  &  15.32  &  14.70  & 14.33 & 14.36 & 14.18 & 13.93 & L/early-T & red star? \\ 
2 & 261.10120 & $+$60.03591 &  15.74  &  15.06  &  14.77  & 13.96 & 13.97 & 14.03 & 14.00 & L/early-T & \\ 
3 & 261.13129 & $+$60.05562 &  16.59  &  16.05  &  15.20  & 14.59 & 14.42 & 14.52 & 14.36 & L/early-T & galaxy? \\ 
\enddata
\end{deluxetable}
\clearpage

\begin{deluxetable}{lllrrrrrrrrrll}
\tabletypesize{\footnotesize}
\rotate
\tablecaption{Shallow Survey brown dwarf candidates\label{tab-FLAMEX}}
\tablewidth{0pt}
\tablehead{\colhead{\#} & \colhead{RA [deg]} & \colhead{Dec [deg]} & \colhead{$B_W$} & \colhead{R} & \colhead{I} & \colhead{J}  & \colhead{K} & \colhead{[3.6]} & \colhead{[4.5]} & \colhead{[5.8]} & \colhead{[8.0]} & \colhead{Type} & \colhead{Notes} }
\startdata
 1 & 216.278002 & $+$34.355660 & \nodata & \nodata &  23.48  &  19.53  &  19.07  &  17.63  &  17.34  & \nodata & \nodata & late-T & red galaxy? \\ 
 2 & 216.603543 & $+$34.140991 & \nodata & \nodata &  23.51  &  20.59  &  19.41  &  18.57  &  17.74  & \nodata & \nodata & late-T & red galaxy? \\ 
 3 & 217.032547 & $+$34.098458 & \nodata & \nodata & \nodata &  20.32  &  19.16  &  18.07  &  17.58  &  16.64  & \nodata & late-T & red galaxy? \\ 
 4 & 217.462015 & $+$33.503213 & \nodata & \nodata &  22.21  &  16.88  &  16.99  &  15.70  &  15.12  &  15.21  &  14.59  & T4.5 & J142950.9$+$333011 \\ 
 5 & 217.786508 & $+$33.139283 & \nodata &  22.75  &  20.40  &  17.38  &  16.30  &  15.80  &  15.69  &  15.35  &  15.12  & L/early-T & \\ 
 6 & 217.920577 & $+$33.295865 &  27.24  &  22.30  &  20.04  &  17.42  &  16.41  &  15.79  &  15.69  &  15.52  &  15.88  & L/early-T & \\ 
 7 & 218.001634 & $+$33.925557 & \nodata &  26.00  &  22.33  &  19.29  &  18.63  &  17.70  &  17.22  &  16.32  & \nodata & late-T & red galaxy? \\ 
 8 & 218.003005 & $+$33.949375 & \nodata &  23.89  &  21.35  &  18.89  &  18.80  &  17.79  &  17.46  &  16.90  &  16.07  & late-T & red galaxy? \\ 
 9 & 218.303091 & $+$34.477201 &  26.43  &  21.69  &  19.23  &  16.33  &  15.20  &  14.58  &  14.69  &  14.47  &  14.42  & L/early-T & \\ 
10 & 218.335896 & $+$33.850797 & \nodata & \nodata & \nodata &  20.25  &  18.86  &  17.12  &  17.01  &  16.40  &  15.21  & L/early-T & bad I photometry? \\ 
\enddata
\end{deluxetable}
\clearpage

\begin{deluxetable}{lccccccccc}
\tablecaption{Companion search efficiency and completeness\label{tab-ec-ns}} 
\tablewidth{0pt}
\tablehead{ 
& \hspace{0.5cm} & \multicolumn{2}{c}{early-T} &
  \hspace{0.5cm} & \multicolumn{2}{c}{late-T} \\
& & \colhead{$D_{th}$} & \colhead{$\cal C \& \cal E$} &
  & \colhead{$D_{th}$} & \colhead{$\cal C \& \cal E$} }
\startdata
$d =  5$~pc && 0.52 & 99.0\% && 0.57 & 97.2\% \\
$d = 10$~pc && 0.49 & 97.3\% && 0.55 & 97.1\% \\
$d = 20$~pc && 0.42 & 89.5\% && 0.52 & 96.3\% \\
\\
\enddata
\end{deluxetable}
\clearpage


\clearpage

\clearpage

\clearpage

\clearpage

\clearpage

\clearpage

\end{document}